\documentclass[prd,a4paper,twocolumn,preprintnumbers,nofootinbib,superscriptaddress]{revtex4}

\pdfoutput=1
\usepackage[english]{babel}
\usepackage{amsmath,amssymb,amsfonts, bm,bbm,slashed, subdepth}
\usepackage{graphicx}
\usepackage{hyperref}
\usepackage{enumerate}
\usepackage{setspace}
\usepackage{booktabs, tabularx}
\usepackage{units}
\usepackage{color}

\newcommand{\be}{\begin{equation}}
\newcommand{\ee}{\end{equation}}
\newcommand{\ba}{\begin{array}}
\newcommand{\ea}{\end{array}}
\newcommand{\bea}{\begin{eqnarray}}
\newcommand{\eea}{\end{eqnarray}}
\newcommand{\balg}{\begin{align}}
\newcommand{\ealg}{\end{align}}
\newcommand{\bit}{\begin{itemize}}
\newcommand{\eit}{\end{itemize}}

\newcommand{\sva}{\langle \sigma_\text{anni} v\rangle}
\newcommand{\ssi}{\sigma_\text{SI}}

\makeindex

\begin{document}
\preprint{ULB-TH/16-13}

\title{

Can the relic density of self-interacting dark matter be due to annihilations\\
into Standard Model particles? 
}

\author{Xiaoyong Chu}
\email{xchu@ictp.it}
\affiliation{International Centre for Theoretical Physics, ICTP\\
Strada Costiera 11, 34014 Trieste, Italy
}

\author{Camilo Garcia-Cely}
\email{Camilo.Alfredo.Garcia.Cely@ulb.ac.be}
\affiliation{Service de Physique Th\'eorique, Universit\'e Libre de Bruxelles, Boulevard du Triomphe, CP225, 1050 Brussels, Belgium}

\author{Thomas Hambye}
\email{thambye@ulb.ac.be}
\affiliation{Service de Physique Th\'eorique, Universit\'e Libre de Bruxelles, Boulevard du Triomphe, CP225, 1050 Brussels, Belgium}

\begin{abstract}

Motivated by the hypothesis that dark matter self-interactions provide a  solution to the small-scale structure formation problems,
we investigate the possibilities that the 
relic density of a self-interacting dark matter candidate 
can proceed from the
thermal freeze-out of annihilations  into Standard Model particles.
We find that scalar and Majorana  dark matter in the mass range of $10-500$ MeV, coupled  to a slightly heavier massive gauge boson, are the only possible candidates in agreement with multiple current experimental constraints. Here dark matter annihilations take place at a much slower rate  than the self-interactions simply because the interaction connecting the Standard Model and the dark matter sectors is small. We also discuss prospects of establishing or excluding these two scenarios in future experiments.  
\end{abstract}

\maketitle

\section{Introduction}

Observations of dynamics of galaxies, clusters of galaxies and the Universe at large scales strongly support the  Cold Dark Matter\,(CDM) paradigm. This suggests that  most of the matter of the Universe consists  of  non-relativistic collisionless particles not present in the Standard Model\,(SM) of particle physics. 
In spite of these successes, within the CDM paradigm, a number of difficulties - such as the \emph{too-big-to-fail}\cite{BoylanKolchin:2011de, BoylanKolchin:2011dk} and the \emph{core-vs-cusp} \cite{Walker:2011zu,deNaray:2011hy} problems- have been found in N-body simulations of formation of small-scale structures, most notably of dwarf and low-surface-brightness galaxies. For a review of these problems, see Ref.~\cite{Weinberg:2013aya}.

Strongly self-interacting dark  matter\,(SIDM) is a plausible solution to  some of these 
challenges~\cite{Spergel:1999mh}.
Its key ingredient is the hypothesis that dark matter\,(DM) particles scatter off each other in small-scale structures   with a cross section per unit of mass of around $\unit[1]{cm^2/g}$~\cite{Wandelt:2000ad,Vogelsberger:2012ku,Rocha:2012jg,Peter:2012jh,Zavala:2012us,Vogelsberger:2014pda,Elbert:2014bma,Kaplinghat:2015aga}. This corresponds to $\unit[10^{12}]{pb}$ for DM masses around $\unit[1]{GeV}$, which is orders of magnitude above  the standard thermal freeze-out cross section of about \unit[1]{pb}; for a comprehensive discussion  of alternative production regimes in the context of SIDM, see Ref.~\cite{Bernal:2015ova}. Clearly, if DM  undergoes a thermal freeze-out in the early Universe,
some mechanism should be at work in order to explain this disparity of cross sections. 

Two of these mechanisms have been discussed extensively in the literature. One of them is invoking a  light mediator enhancing DM self-interactions via non-perturbative effects in small-scale structures~\cite{Feng:2009hw, Buckley:2009in, Loeb:2010gj, Tulin:2013teo}. The other one is considering DM annihilation processes in the early Universe that are induced by a relatively strong interaction but that are nevertheless phase-space suppressed due to the presence of many particles in their initial state~\cite{Carlson:1992fn, Hochberg:2014dra}. In both cases, the production of DM proceeds via annihilations within the hidden sector. In the first case, DM annihilates dominantly into the light mediator responsible for the self-interactions. In the second case, three or four~\cite{Bernal:2015xba} DM particles annihilate into two of them. Thus in both scenarios an interaction connecting the DM and the SM sectors is not necessary for the DM self-interactions or annihilations\footnote{Such an interaction might be nevertheless necessary for other concerns. For instance, for inducing the decay of the mediator in order to satisfy BBN and CMB constraints \cite{Kaplinghat:2013yxa} or for establishing kinetic equilibrium between DM and SM thermal bath in the 3-to-2 framework \cite{Hochberg:2014dra,Bernal:2015bla,Lee:2015gsa}.}.  Needless to say, without such an interaction we will never discover the DM particle 
and will only be able to probe it through its gravitational and/or its self-interaction effects.

Although it is possible that after inflation no connector between both sectors has played any 
major role for DM annihilation and self-interaction processes, in this work we explore 
a third mechanism, largely overlooked to the best of our knowledge, in which the relic density of SIDM stems from
the freeze-out of its annihilations into SM particles.
In other words, 
we will show that  DM  self-interacting in a hidden sector must not necessarily annihilate into particles beyond the SM.

If, as we will assume all along this work, the large self-interaction cross section does not result from non-perturbative effects associated to the exchange of a lighter mediator, the DM particle must  lie below the GeV scale.
Searches of particles beyond the SM severely constrain such scenarios, basically restricting sub-GeV DM to be a singlet under the SM gauge group and requiring it to have rather small  interactions with the SM particles. This is the mechanism we explore in this work: even though the DM sector has relatively strong interactions, its portal to SM particles - which are the dominant annihilation products of DM- is comparatively small and  leads to a thermal freeze-out in agreement with the observed abundance of DM. 

This article is organized as follows. We start off in Sec.~\ref{sec:SIDMtoSM} by determining the possible scenarios giving rise to annihilations of sub-GeV DM into SM particles, based on the four possible portal interactions that are allowed by  SM symmetries. From this discussion, only one scenario emerges, which is based on
the portals that include an extra gauge boson. In Sec.~\ref{sec:VectorPortal}, we discuss such scenario in detail and examine the long list of corresponding experimental and observational constraints.
Possibilities of future particle physics tests, associated to the fact that DM annihilates into SM particles, are also analyzed.  
Finally, we present our conclusions in Sec.~\ref{sec:conclusions}.

\section{Four portals for SIDM annihilations into SM particles }
\label{sec:SIDMtoSM}

\subsection{Basic requirements}

 In order that SIDM annihilates dominantly into SM particles, there is a number of preliminary basic requirements that it must fulfill. These are:

\begin{itemize}
\item  {\bf The DM mass must be below the GeV scale}.

As already mentioned above, we do not consider the possibility of a mediator with mass much smaller than  the DM mass $m_\text{DM}$, inducing large DM self-interactions through non-perturbative effects.
This is because,  if the mediator is a particle beyond the SM,  such an option would easily allow the DM particles to annihilate dominantly into a pair of mediators, rather than into SM particles. If instead the mediator inducing non-perturbative effects is a SM particle such as a photon or a massive boson, a sufficiently  large self-interaction cross section could hardly be accommodated without violating experimental constraints~\cite{Feinberg:1968zz,Tulin:2013teo,McDermott:2010pa}.

In absence of lighter mediators, provided the associated dark sector couplings, $g_D$, have perturbative values,
the self-interaction cross section can be calculated 
by means of the ordinary Born expansion in the small-velocity limit\cite{Tulin:2013teo}.  
In this case, for a self-interaction induced by the exchange of a mediator with mass of order $m_\text{DM}$, dimensional analysis shows that $\ssi/m_\text{DM} \sim \alpha_D^2 /m_\text{DM}^3$, with $\alpha_D=g_D^2/4\pi$. Taking $\alpha_D\lesssim \mathcal{O}(1)$, this implies that the DM mass must lie roughly below $\unit[500]{MeV}$. This bound is much lower if the self-interaction mediator is much heavier than the DM particle.
\item \textbf{Kinematically allowed DM annihilation channels}.
 For such a low mass range, DM must necessarily annihilate into one of the few kinematically allowed SM channels:
 $\text{DM}\,\text{DM}\rightarrow \nu \bar{\nu}, \nu \nu, e^+ e^-, \mu^+\mu^-,\gamma \gamma$ and $\text{DM}\,\text{DM}\rightarrow u\bar{u}, d \bar{d}$ (i.e.~$\text{DM}\,\text{DM}\rightarrow \pi^+\pi^-, \pi^0 \pi^0$ at the scale under consideration). 
\item \textbf{DM must be a gauge singlet.} This is a consequence of the first requirement. For instance, particles with $SU(2)_L\times U(1)_Y$ quantum numbers and a mass well below the electroweak scale would have been seen in the
decay 
of the $Z$ boson at LEP~\cite{Cao:2007rm}. There are exceptions to this rule, but they  entail some degree of fine tuning, so  we will not consider them. 
 An example would have been DM as the CP-even neutral component of a scalar doublet. It can be  light and still escape the bound coming from the  width of the $Z$ boson if the CP-odd component  in the doublet has a mass above $m_Z$.
\item  \textbf{Extra particles mediating the annihilation are singlets.} For the same reason, any additional non-singlet particle  mediating DM annihilations into SM particles  would have to be   much heavier than the DM particle, typically above $m_Z/2$ or higher.
This would suppress the annihilation cross section by powers of this high mass. As a result in this case we find that the thermal freeze-out could only be obtained for couplings on the verge of non-perturbativity.
 
In order to illustrate this, let us consider the tree-level annihilation of a Dirac DM particle into neutrinos via the exchange of the neutral component of a $SU(2)_L$ doublet in the t-channel.
The annihilation is suppressed by four powers of the mass of the exchanged particle.  Concretely,  one obtains an annihilation cross section into neutrinos of a given flavor equal to 
\be
\langle \sigma_\text{anni} v \rangle=\frac{y^4}{32\pi}\frac{m^2_\text{DM}}{m^4_\phi} \,,
\label{sigmavnuLnuL}
\ee
where $m_\phi$ is the mass of the neutral scalar in the t-channel and $y$ is the Yukawa coupling in the interaction ${\cal L} = y \,\overline {L_L} \phi \,\psi_{DM}$.
 For sub-GeV DM and $m_\phi$ of order $m_Z$, this gives the thermal freeze-out cross section only for quite large Yukawa couplings, at the verge of non-perturbativity\footnote{Note that for  Majorana or scalar DM, the exchange of a doublet in the t-channel also induces annihilations into neutrinos. However, in those cases, the cross section is even more suppressed than for  Dirac DM, because it is proportional to the neutrino masses (See e.g.~\cite{Garny:2015wea,Lindner:2010rr}). In fact, we did not find any viable scenario where DM  annihilations into SM fermions are suppressed by only two powers of the mass of the exchanged particle.}, namely, $y \gtrsim  5.6 \cdot (\unit[100]{MeV}/m_\text{DM})^{1/2}\cdot (m_\phi/\unit[100]{GeV})\,$.  Or in other words, imposing $y \lesssim \sqrt{4\pi}$ leads to $m_\text{DM}\gtrsim \unit[200]{MeV}\cdot (m_\phi/\unit[100]{GeV})^2$. Also, notice that the same type of Yukawa interactions potentially leads to DM annihilations into charged leptons, and that a thermal rate for that channel is forbidden by indirect detection constraints, as discussed below. As a result of all these, we will not consider any further such kinds of contrived scenarios.

  Note that SIDM annihilations into photons are suppressed not only by the loop factor but also, in a similar way, by the large mass of the charged mediator in the loop. The same remarks apply to other processes leading to sharp spectral features, such as virtual internal bremsstrahlung (since they require a charged mediator in the t-channel).
 
\end{itemize}

The previous four criteria greatly simplify the discussion and highly limit  the number of scenarios where SIDM could freeze out from annihilations into SM particles, as we will see in the following.

\subsection{Four portals to the SM}

In a renormalizable theory, if both the DM and the particle mediating the annihilation process are singlets, they can  only communicate with the SM particles via the so-called portals. They correspond to the four possible ways of 
 building, out of SM fields, a gauge singlet operator of dimension less than four~\cite{Patt:2006fw, Essig:2013lka}, namely
\begin{eqnarray}
&&\text{Vector portal}:  \quad  \,\,\overline{\psi_\text{SM}}\gamma^\mu \psi_\text{SM}\nonumber\\
&&\text{Kinetic portal}:  \quad  F^{\mu\nu}_Y \nonumber\\
&&\text{Higgs portal}: \quad \,\,\,\, H^\dagger H\nonumber\\
&&\text{Neutrino portal}: \,\bar{L} H\nonumber
\end{eqnarray}
where $\psi_\text{SM}$ is any SM fermionic chiral multiplet, $F^{\mu\nu}_Y$ is the hypercharge field strength, $H$ is the SM scalar doublet and $L$ is one of the lepton doublets. 

On the one hand, the fermion bilinear can only be coupled in a renormalizable way to a vector boson field, $Z'_\mu$. On the other hand,
the hypercharge field strength can only couple to the field strength of a vector boson,  $Z'_{\mu\nu}$, through a kinetic mixing interaction
\be
{\cal L } = -\frac{\kappa}{2} F^{\mu\nu}_Y Z'_{\mu\nu}.
\label{eq:LZp}
\ee
Thus, from the exchange of a $Z'$, both sectors can communicate through either of these two portals or through both.

As for the  $H^\dagger H$ bilinear, it can couple to any single scalar operator with dimension two. The most general form is 
\be
{\cal L}=  H^\dagger H\cdot  (\mu_i\phi_i+\lambda_{ij} \phi_i \phi_j)\,
\label{eq:LHiggs}
\ee
where $\phi_i$ are singlet scalar fields. Finally, the bilinears $\bar{L} H\nonumber$  must couple 
to fermion singlets, i.e.~to right-handed neutrinos
\be
{\cal L} =   y_\alpha \overline{L_\alpha} H \nu_R + h.c.\,,
\label{eq:nuR}
\ee

All these portals induce annihilations at tree level. In principle, such annihilations can proceed in three ways: 
from a s- or a t-channel exchange and from a quartic bosonic interaction.
We now discuss each case separately:
\begin{enumerate}
\item \textbf{Tree-level annihilation via a $Z'$ exchange}.  
Since the $Z'$ couples to a pair of SM particles, the annihilation necessarily takes place through a s-channel exchange, either from the vector portal, or from the kinetic portal, or from both.
Furthermore, DM naturally self-interacts at an unsuppressed rate via the exchange of the $Z'$ boson. 
Hence, this scenario is particularly attractive and minimal. We will discuss it in detail in Sec.~\ref{sec:VectorPortal}. It differs from  previous SIDM studies involving  $Z'$ bosons by the fact that here DM is  lighter than such particles, and thus  does not annihilate  into a pair of them but into SM particles through the s-channel exchange of a $Z'$.  Notice that models with MeV  DM coupled to a heavier $Z'$ boson  have also been considered in contexts different from SIDM (See e.g.~\cite{Huh:2007zw,Pospelov:2007mp} for its implications on the galactic 511 keV line). 

\item \textbf{Tree-level annihilation via the Higgs portal}. 
If a field $\phi$ entering in the Higgs portal above has no linear interactions -in particular no vacuum expectation value $\langle \phi \rangle$ and no term  $ \mu \,\phi \,H^\dagger H $
in the Lagrangian- it can  only communicate to the SM through the interaction ${\cal L}= \lambda\, \phi \phi\, H^\dagger H $. 

In this case, this field could be a DM candidate and annihilate through a Higgs boson exchange into light SM particles. However, taking into consideration that (i) the Higgs boson can not decay into DM with a large rate (in order to avoid the LHC bound on its invisible decay  width), (ii) the Yukawa couplings of light fermions are very small,  and (iii) the Higgs boson  is much heavier than the sub-GeV DM candidate discussed here, 
we conclude the exchange of a Higgs boson can not mediate annihilations processes fast enough in order to lead to the observed relic density.

Instead,  if  there is a Higgs portal interaction linear in $\phi$ (i.e. $\mu\neq 0$, or $\lambda\neq0$ when $\langle \phi \rangle\neq0$),
after electroweak symmetry breaking, the scalar field $\phi$ mixes with the SM model scalar and inherits its Yukawa couplings to ordinary fermions. It is thus unstable and we need an additional particle as DM candidate, which annihilates into SM fermions via the scalar portal.  The tree-level annihilations  of such candidate can take place  via the exchange of the two scalar mass eigenstates in the s-channel: the Higgs boson and the other scalar arising from the mixing.  The former case is excluded in the (i)-(iii) way above. The latter case is also excluded because, even if the other scalar arising from the mixing is lighter, the corresponding annihilations turn out to be still quite suppressed because its interactions
are still proportional to the small Higgs Yukawa couplings. Thus, unless we sit  on the $m_\phi \simeq 2 m_{DM}$ resonance to enhance the annihilation process, the relic density cannot be accounted  via the freeze-out mechanism and this scenario is therefore not viable. 

\item \textbf{Tree-level exchange via the neutrino portal}. The neutrino portal requires one or more right-handed neutrinos. Since it necessarily induces a mixing of these particles with the SM neutrinos, this portal offers the possibility of having DM annihilations into active neutrinos.
At tree level, for scalar DM as well as fermion DM, there are three ways to induce such an annihilation, two in the s-channel (via the exchange of a scalar or a vector particle)  and one in the t-channel. In all these cases, this requires the existence of an extra particle in addition to the DM and singlet neutrinos. However, the resulting 
neutrino mixing is highly bounded from above by neutrino-mass constraints, and the corresponding
annihilation cross section turns out to be too small.
Thus, the neutrino portal does not work for our purposes either. 
\end{enumerate}

\subsection{Surpassing indirect detection constraints}

 DM annihilations into SM particles can potentially produce a significant flux of cosmic rays,  specially  if they are produced in astrophysical systems where the DM concentration is known to be very high (see e.g.~Ref.~\cite{Bertone:2004pz}). Likewise,  such annihilations also lead to distortions of the CMB spectrum \cite{Lopez-Honorez:2013lcm, Steigman:2015hda,Slatyer:2015jla} or to a departure from the predictions of standard Big Bang Nucleosynthesis\,(BBN)~\cite{Henning:2012rm,Jedamzik:2009uy}. The non-observation of these phenomena leads to stringent bounds on annihilations cross sections, specially for sub-GeV DM.  In fact, one finds that  an annihilation cross section into SM particles around the thermal freeze-out value is excluded for such masses, except in two cases:
\begin{itemize}
\item {\it If DM annihilates almost exclusively into neutrinos}~\cite{Frankiewicz:2015zma}. 
The neutrino portal would have been interesting in this respect because it gives rise to such situation naturally. Nonetheless, it does not work in the context of SIDM, as mentioned above. 
\item {\it If DM annihilations are velocity-suppressed}. In this case all fluxes are suppressed because in all the systems from which the bounds are derived,  DM moves with very small velocities compared to the freeze-out epoch~\cite{Goldberg:1983nd, Bergstrom:2004cy, Bringmann:2007nk, Barger:2009xe, Giacchino:2013bta}. 
More quantitatively, for velocity-suppressed DM annihilations, the cross section can be expanded as $\sigma_\text{anni} v = b v^2$, where $v$ is the relative DM velocity at a given epoch.  The observed DM density $\Omega h^2$ fixes the quantity $b$. Using the  instantaneous freeze-out approximation as reported in Ref.~\cite{Griest:1990kh}, one gets the following
relic abundance
\begin{equation}
\Omega h^2 = \left(\frac{\unit[1.07 \times 10^9]{GeV^{-2}} }{3 \,g_*(x_f)^{1/2} \, M_\text{Pl}\, b }  \right) n\, x_f^2\,,
\label{eq:Omegah2}
\end{equation}
where $x_f \approx 20$ is the usual inverse freeze-out temperature,  $n=1$ for self-conjugate DM and $n=2$ in the opposite case.  
Taking the relic density equal to $\Omega h^2 = 0.1199\pm 0.0027$~\cite{Ade:2013zuv}, the previous procedure leads to values of $b$ of about $10^{-25}$\,cm$^3/$s for sub-GeV DM. For such values, scenarios with velocity-suppressed DM annihilations are not constrained by indirect searches. Both photo-dissociation of $^4$He and photon-decoupling processes happen when the DM particles are already highly non-relativistic. Therefore,  CMB and BBN bounds are irrelevant here and the most stringent constraints  can only come from DM indirect searches in dark halos at very low redshifts. However, even there, current experiments give upper bounds on $ b v^2 $ of around $10^{-28}$-$10^{-27}$\,cm$^3/$s~\cite{Boddy:2015efa} (with $v\lesssim 10^{-2}$ in dark halos, as given by cosmological simulations), leaving velocity-suppressed annihilations cross sections unconstrained.
\end{itemize}

In practice, a velocity suppression in the annihilation process means that the s-wave piece of the corresponding cross section is not allowed.  For the portals which have been found to be viable above, i.e.~the vector and kinetic portals,
this is only possible in specific cases. To see that, suppose that DM annihilation takes place via the s-wave, i.e.~with orbital angular momentum $L=0$. In order to exchange a $Z'$, we must have a state with total angular momentum $J=1$, or equivalently total spin $S=1$. This  is not possible for scalar or Majorana DM since they both lead to $S=0$; the state $S=1$ is symmetric for a pair of fermions in the $L=0$ configuration and is thus banned for Majorana particles. As a result, if we couple the $Z'$ boson to scalar or Majorana DM particle, we naturally obtain velocity-suppressed annihilations and evade indirect detection bounds. 

Note that both of these scenarios can hardly be probed by high-energy colliders. For example,  
missing-energy searches at LHC are able to exclude thermal freeze-out mechanism for $m_{DM}\lesssim O(10)$\,GeV if $m_{Z'}\gtrsim 50\,$GeV\,\cite{Fox:2011fx}. But  if the mediator is also  light, thermal freeze-out of DM only requires much weaker couplings with SM particles, which is well beyond the reach of high-energy collider experiments. This has been shown in various so-called simplified model studies (for a recent analysis, see \cite{Bell:2015rdw}). In contrast, it is well known that data-intensive experiments at relatively low energy, as well as other precision measurements, may provide very strong bounds for $Z'$-portal models at the scale below GeV~\cite{Essig:2013lka}. This has been discussed in the previous literature~\cite{Izaguirre:2015yja, Izaguirre:2015pva} and will be investigated  in the SIDM framework for both Majorana and scalar DM in the next Section.

 In conclusion, the previous natural list of constraints and criteria
point towards a unique scenario with two variants: Majorana or scalar DM annihilating into light  SM leptons or quarks through a heavier spin-1 particle exchange.

\section{Scenarios with  a $Z'$ boson }
\label{sec:VectorPortal}

As said above, a $Z'$ can be exchanged between the DM particle and the SM sector from the vector portal, the kinetic mixing portal, or both. In either case, we assume the $Z'$ to be associated to a $U(1)_D$ gauge interaction  with a mass originating from the Brout-Englert-Higgs  or the Stueckelberg mechanisms.

If some of the light SM particles are charged under the $U(1)_D$ group and if in addition there is no kinetic mixing, DM communicates with the SM sector only through the vector portal. Provided the $U(1)_D$ gauge coupling and the corresponding charges are of order one (as it is the case for the known gauge groups), this possibility is highly constrained by collider experiments. In particular, the bound $m_{Z'}>2.1$~TeV holds if the $Z'$ sizably couples to SM leptons~\cite{Alves:2015pea}. For a leptophobic  $Z'$, the bound is weaker, but in general still requires $m_{Z'}$ heavier than few hundred GeVs, depending on its exact couplings to quarks~\cite{Alitti:1993pn, Erler:2009jh, Alves:2013tqa}. Such heavy $Z'$ can not induce DM  annihilations with perturbative couplings, thus we will not consider this portal any  further (although it could certainly work in special cases where the relevant gauge couplings  have sufficiently small values).

In the following, we will consider the opposite option, where all SM particles have no $U(1)_D$ charges, but where there is a non-zero kinetic mixing interaction, as given in Eq.~\eqref{eq:LZp}, so that the communication of both sectors solely occurs through this portal.

This leads to a
highly predictive and minimal scenario, in which all $Z'$ couplings to SM particles are known up to the overall multiplicative kinetic mixing parameter. Concretely,
 after electroweak symmetry breaking, Eq.~\eqref{eq:LZp} gives rise to the following $Z'$ interactions
\be
{\mathcal L} \supset g_D J_\text{DM}^\mu Z'_\mu + \epsilon e \bar J_\text{EM}^\mu Z'_\mu + \epsilon' g_Z \bar J_\text{Z}^\mu Z'_\mu ,
\ee
where $J_\text{DM}^\mu $, $J_\text{EM}^\mu $ and $J_\text{Z}^\mu $ are the dark, the QED and the weak neutral currents. The exact expression of $J_\text{DM}^\mu $ depends on the dark matter spin, and  will be given separately for each case below.  Also, the vector boson couplings to the SM currents are given, to leading order in the kinetic mixing parameter, by~\cite{Babu:1997st}
\begin{align}
\epsilon& \approx \cos{\theta_W}\, \kappa\,, & \epsilon' \approx \frac{m^2_{Z'}}{-m^2_Z+m^2_{Z'}}\sin{\theta_W} \kappa \,,
\end{align}
where $\theta_W$  is the Weinberg angle. 
It follows that $\epsilon \gg \epsilon'$ for $m_Z \gg m_{Z'}$. We can thus safely neglect the interaction term  involving $J_\text{Z}^\mu$. This in turn means that annihilations into neutrinos are negligible and only final states including light charged leptons or quarks can be responsible for the DM freeze-out. Furthermore, this implies that the DM particle must be heavier than the
 electron.

\subsection{Majorana dark matter}

\begin{figure*}[t]
\centering\includegraphics[width=3.2in]{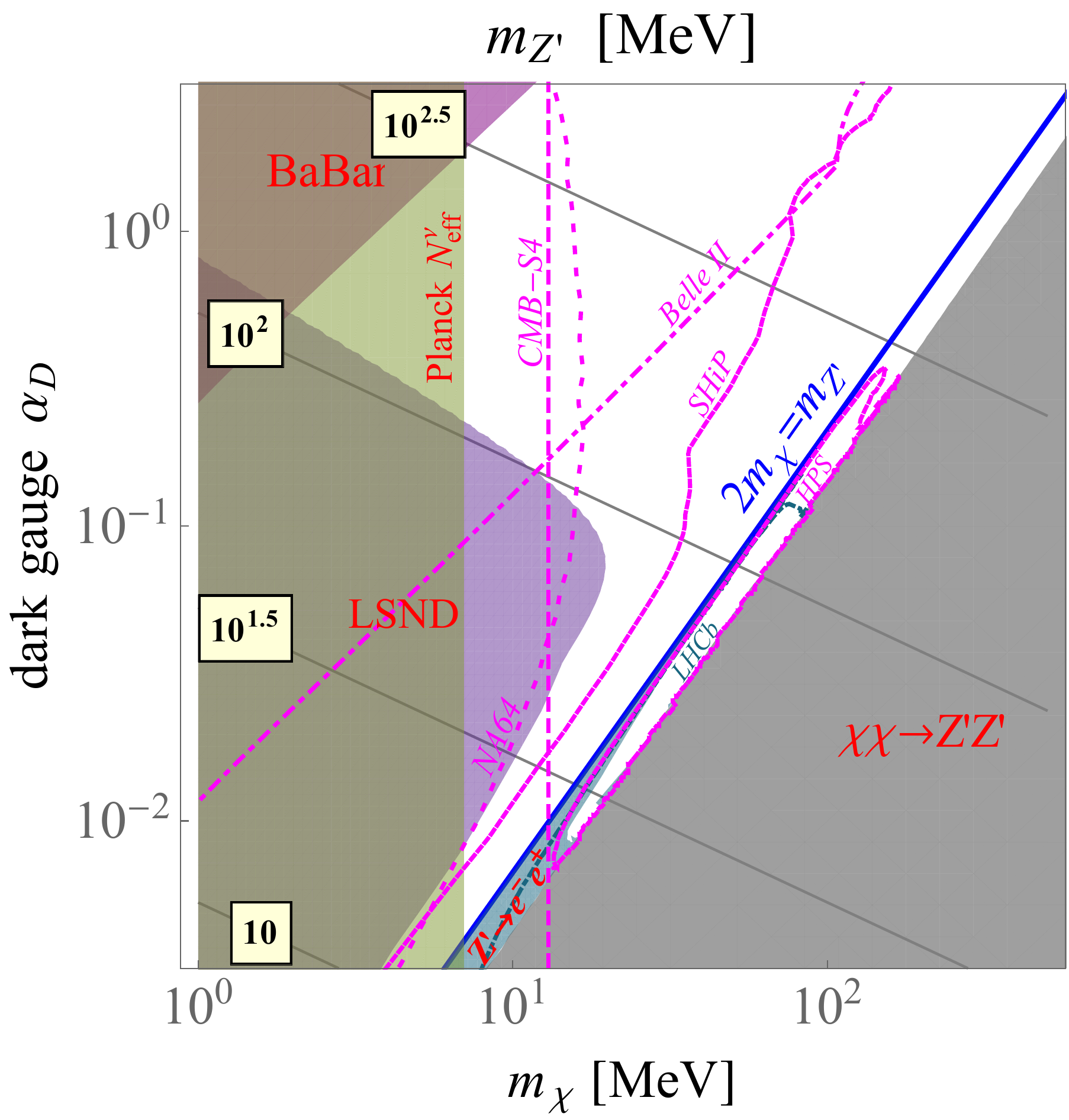}
\centering\includegraphics[width=3.2in]{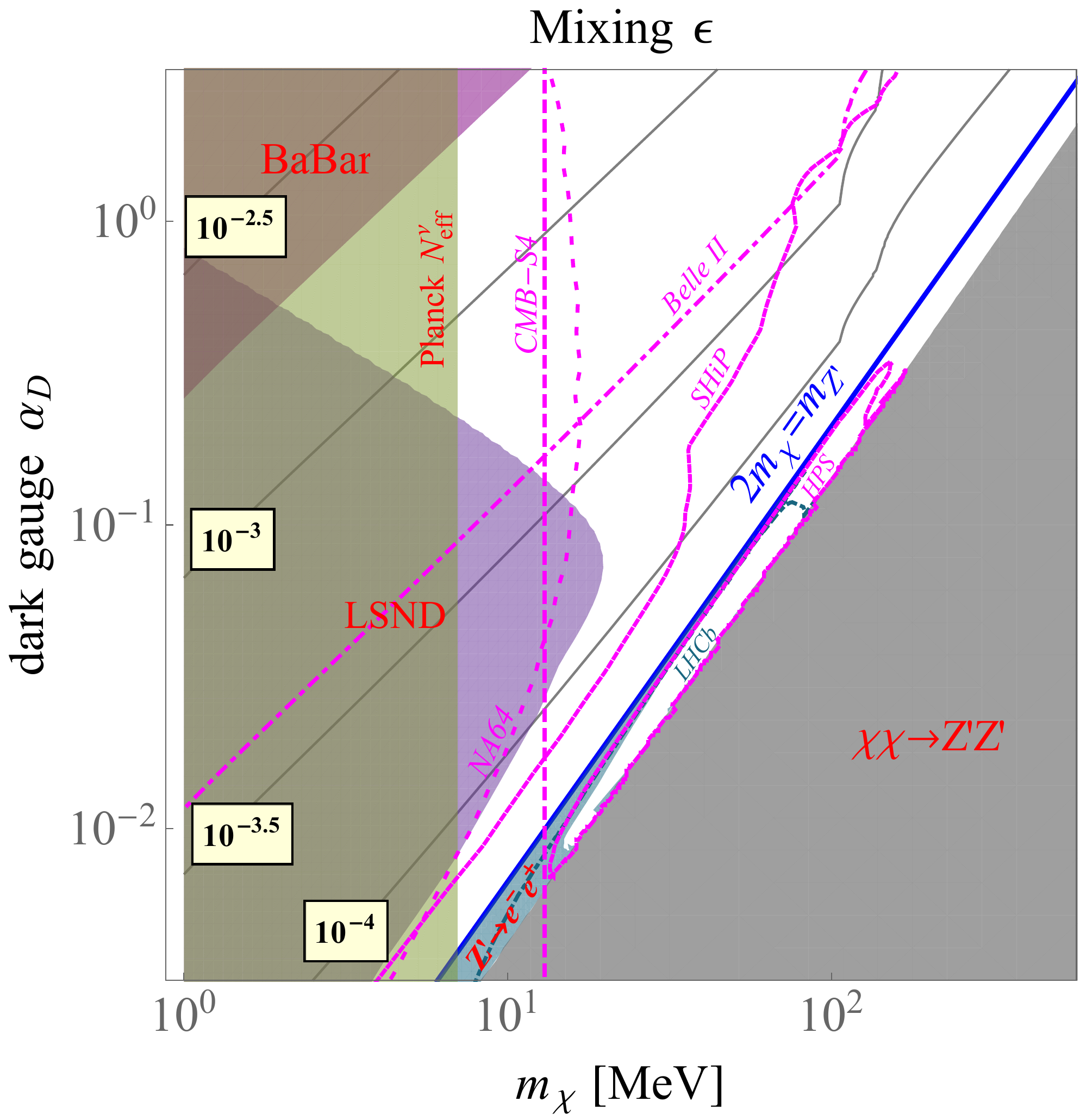}
\vskip -.5cm
\caption{{\it $Z'$ portal for Majorana DM}.  As a function of the DM mass $m_\chi$ and dark coupling $\alpha_D$, the solid contour lines show the values of $Z'$ mass (\textit{left}) and kinetic mixing parameter $\epsilon$ (\textit{right}) satisfying the relic density and the self-interaction constraints, as given in Eqs.~\eqref{eq:Omegah2} and \eqref{Zportal:MajoranaSI}.  All shaded regions are experimentally excluded in various ways (see text for details). In the shaded region at the \textit{right-bottom} corner, the dark annihilation $\chi\chi\to Z'Z'$ is too fast to account for the DM abundance. Non-solid  (colored) lines show the expected sensitivities of future experiments. }
\label{fermion:Zportal}
\end{figure*}

If DM is made of a Majorana fermion $\chi$, their  current coupling it to the $Z'$ is given by $J_\text{DM}^\mu = \bar \chi \gamma^\mu \gamma^5 \chi $.  
For the DM mass range of interest, the annihilation channels are 
$\chi\chi \to Z'^* \to {\bar f} f$,
with $f$ an electron, a muon, an up quark or a down quark (i.e.~pions for the last two cases). For a given fermion of electric charge $q_f$ and color $N_f$, the annihilations cross section is given by
\be
\langle \sigma_\text{anni} v \rangle  \simeq \frac{16\pi \epsilon^2\alpha \alpha_D  \sum_{f} \,N_f\, q_f^2 \,   (1-r_f)^{1/2} (2+r_f)\,v^2}{3 m_{\chi}^2\left((r_{Z'} - 4)^2 + r_{Z'}^2\Gamma_{Z'}^2/{m^2_{Z'}}\right)
}  \,,
\label{Zportal:MajoranaFO}
\ee
with $r_a = m_a^2/m_\chi^2$. Summing over all kinematically allowed channel, the relic density is given by Eq.~\eqref{eq:Omegah2}. We work under  the approximation that, for $m_\chi>m_{\pi}$, terms in Eq.~(\ref{Zportal:MajoranaFO}) associated to  the up and down quarks give the inclusive cross section into pions.

As for the self-interaction hypothesis, we have
\be
\ssi/m_\chi \simeq 512\, \pi \alpha_D^2 m_\chi /m_{Z'}^4 \simeq \unit[1]{cm^2/g}\,, \label{Zportal:MajoranaSI}
\ee
where the low velocity limit has been taken. Note that the non-observation of an offset between the mass distribution of DM and galaxies in the Bullet Cluster has been claimed to constrain the self-interacting cross section, $\ssi /m_\text{DM}<1.25$ cm$^2$/g at $68\%$~CL~\cite{Clowe:2003tk,Markevitch:2003at,Randall:2007ph}. However, recent simulations suggest that stronger self-interactions are still allowed~\cite{Robertson:2016xjh, Elbert:2014bma}. In the following, we will always take $\unit[1]{cm^2/g}$ as a benchmark value for $\ssi /m_\text{DM}$. Modifying it by a factor of a few would only affect our conclusions mildly.

The relic density and self-interactions constraints just mentioned fix $m_{Z'}$ and $\epsilon$ as functions of DM mass $m_\chi$ and  dark fine structure constant $\alpha_D$.
Fig.~\ref{fermion:Zportal} shows the values we obtain by following this procedure.

As said above, to prevent a fast $\text{DM}\text{DM} \rightarrow Z' Z'$ annihilation leading to a too suppressed relic density, one assumes $m_{Z'}\gtrsim m_\text{DM}$. More exactly, this requirement rather leads to $m_{Z'}\gtrsim 1.6m_\text{DM}$ after taking into account Eq.~\eqref{Zportal:MajoranaSI}. All values not satisfying this requirement are shaded in grey in Fig.~\ref{fermion:Zportal}.
For a precise value of $m_{Z'}/m_\text{DM}$ around 1.6, the annihilation rate is Boltzmann suppressed  ($\propto e^{-2m_{Z'}/T}$) just enough to lead to the observed relic density. In this case, both
self-interactions and annihilations constraints are accounted for by the hidden sector interaction and the condition on the connector $\epsilon$ is that it has to be small enough to play only a subleading role in the annihilation process. This way of accounting for both constraints in the hidden sector by means of a threshold effect\cite{Griest:1990kh} has been proposed in Ref.~\cite{D'Agnolo:2015koa} and is operative here. For smaller (larger) values of $m_{Z'}/m_\text{DM}$,  the $\text{DM} \text{DM}\rightarrow Z' Z'$ annihilation rate is very quickly far too fast (slow) to account for the relic density. In the later case, $m_{Z'}/m_\text{DM}\gtrsim 1.6$, the connector interaction can nevertheless account for it. This is the scenario we consider, which leads to a whole allowed region in Fig.~\ref{fermion:Zportal}.

Not surprisingly, Fig.~\ref{fermion:Zportal} reflects the disparity of both cross sections.
On the one hand, the self-interaction constraint requires
a cross section, Eq.~(\ref{Zportal:MajoranaSI}), which is not suppressed in any way, i.e.~a relatively large value of $\alpha_D$ and a relatively light mediator, $m_{Z'}\lesssim~\mathcal{O}(1)$~GeV. On the other hand, for sub-GeV DM the relic density constraint, Eq.~(\ref{Zportal:MajoranaFO}), requires 
a suppression of the annihilation cross section. This can only arise from a suppressed portal, i.e.~mixing parameter $\epsilon$. This way of decoupling both cross sections is an easy way to account for the big difference between them: the annihilation rate is  naturally suppressed with respect to  the self-interactions because the portal, which enters only in the annihilation cross section, is very small. Note that, as Fig.~\ref{fermion:Zportal} shows, for $m_{Z'}\sim 2 m_\chi$ the annihilation cross section displays a resonance, requiring even smaller values of $\epsilon$. 

There is a long list of constraints applying to this scenario. The most relevant ones are shown on Fig.~\ref{fermion:Zportal}. These are:
 
\begin{itemize}
\item {\textbf{Invisible decay of $Z'$ in low energy experiments}}. 
Due to the large value of $\alpha_D$ and small value of $\epsilon$,
when $m_{Z'} \gtrsim 2m_\chi$ (above the blue lines in Fig.~\ref{fermion:Zportal}), the $Z'$ decays invisibly with a branching ratio close to one. Hence, at colliders, the $Z'$ cannot be seen directly and the best way to detect it is from the observation of initial state radiation and missing energy. Note that the cross section for such a signal depends only on $m_{Z'}$ and on the size of the couplings between the $Z'$ and the SM particles in the initial state, i.e. on the size of the $\epsilon$ parameter.
The BaBar collaboration  -searching for the decay of $\Upsilon$(3S) to mono-photon and invisible particles- has  constrained the coupling between $Z'$ and SM particles for $m_{Z'} \lesssim 7.2$\,GeV~\cite{Aubert:2008as, Essig:2013vha}. This constraint is shown as a shaded region in Fig.~\ref{fermion:Zportal}. In addition, the Belle II experiment, which should start taking data after 2018~\cite{Wang:2015kmm}, has the potential to improve the constraint on $\epsilon$ by about one order of magnitude. In Fig.~\ref{fermion:Zportal}, the dot-dashed line in the left-top corner
shows the corresponding projected sensitivity, adapted from the mono-photon search done in Ref.~\cite{Essig:2013vha}. 

Likewise,  by studying the process $e\,Z \to e\,Z Z'$, the fixed-target experiment NA64 at CERN SPS will be able to probe dark photon decays into invisible particles~\cite{Gninenko:2016kpg}. The corresponding projected sensitivity (associated to $10^{11}$ incident electrons) is shown in Fig.~\ref{fermion:Zportal}.

Interestingly, events associated to the  invisible decay of a $Z'$ boson may be recorded in  neutrino experiments as well.  Concretely, if $m_{Z'}<m_{\pi^\pm}$,  depending on its couplings to SM fermions,  the $Z'$ boson might be produced in pion decays and quickly disintegrate into invisible particles, i.e DM in our scenario. In turn, the DM particles might  collide against the electron target, leading to detectable scattering events, similar to the ones induced by neutrinos. Thus, the observed number of such events can be used to constrain our scenario if the $Z'$ is lighter than the pion(s). Using this, the LSND data provides the strongest constraint for very light dark matter~\cite{Batell:2009di}.  This is shown by the shaded region labeled as ``LSND'' in  Fig.~\ref{fermion:Zportal}.  Moreover, using the same search strategy, the SHiP experiment will improve this constraint as shown in Fig.~\ref{fermion:Zportal}. The projected sensitivity in the plot corresponds to a yield of 10 electron-scattering events~\cite{Alekhin:2015byh}.

Also, it has been pointed out that the experiment E137 performed at SLAC three decades ago provides similar constraints on dark photons~\cite{Batell:2014mga}. In  most of the parameter region of our interest, they are less stringent than those of LSND and are thus not shown in our figure.

\item {\textbf{Precision test bounds}}.   When $m_{Z'} \gtrsim 2m_\chi$, this scenario  is also constrained by missing $E_T$ searches at higher energy accelerators, such as LEP and LHC. The current bound is $\epsilon \lesssim 0.23$ for $m_{Z'}$ below few GeV~\cite{Essig:2013vha}.  As $Z'$ mixes with the SM $Z$ boson,  $Z$-pole precision measurements also constrain the mixing parameter, giving $\epsilon \lesssim 0.3$~\cite{Curtin:2014cca}.  Moreover, anomalous magnetic moment measurements of the electron and the muon, as well as neutron-nucleus scattering measurements\,\cite{Barbieri:1975xy}, also lead to upper bounds on $\epsilon$.  Nevertheless, all these precision bounds are looser than other constraints and
we do not show them on Fig.~\ref{fermion:Zportal}.

\item {\textbf{Visible decay of $Z'$}}.   For the  case $m_{Z'} \lesssim 2m_\chi$ (below the blue lines in Fig.~\ref{fermion:Zportal}), the $Z'$ boson decays into pairs of SM fermions. This possibility is extensively considered in experiments looking for dark photons. For instance,  searching for the process $\pi^0\to \gamma Z'$  followed by $Z'\to e^+e^-$,  the NA48/2 collaboration has excluded  $\epsilon \gtrsim  8\times 10^{-7}$ at $90\%$ CL when the ${Z'}$ boson is lighter than the neutral pion~\cite{Batley:2015lha, Feng:2016jff}. We refer  to Ref.~\cite{Essig:2013lka} for a recent review on this constraint and others from beam-dump/collider experiments. This leads to the exclusion of the shaded region labeled as ``$Z'\to e^-e^+$'' in Fig.~\ref{fermion:Zportal}. One can see that there still exists a large  unconstrained region between ${\mathcal O}(10^{-5}) \lesssim \epsilon \lesssim {\mathcal O}(10^{-2})$  for DM masses of few tens of MeVs. Independently of self-interaction constraints, this feature is also shown in Fig.~6 of the review \cite
{Essig:2013lka}. 

Fig.~\ref{fermion:Zportal} also shows the sensitivities expected to be reached in the future by various experiments:  from the proposed Heavy Photon Search (HPS)~\cite{Celentano:2014wya}, looking for leptonic decays of a dark $Z'$ boson, and from the dark photon search at the run 3 of LHCb using charm meson decays~\cite{Ilten:2015hya}. Clearly, these experiments offer real prospects to probe our scenario.
 
\item {\textbf{Cosmological bounds}}.  As DM annihilation is velocity suppressed, it does not directly change BBN predictions or the CMB spectrum, as explained above. It can nevertheless have an indirect effect
from the fact that -after neutrino decoupling at about $1.5$\,MeV~\cite{Fornengo:1997wa}-  late annihilations of DM  into electron-positron pairs, may reheat the thermal bath of photons with respect to the cosmic neutrino bath. This leads to a relatively  colder neutrino sector at the recombination time. Taking $N^\nu_\text{eff} \gtrsim 2.9$ from Planck, we obtain that $m_\chi \gtrsim 7 $\,MeV, as shown by the left shaded region ``Planck'' in Fig.~\ref{fermion:Zportal}~\cite{Heo:2015kra}. Note that assuming an earlier neutrino decoupling would lead to a stronger bound. Proposed CMB precision experiments, referred to as ``CMB-S4'', intend to reduce the uncertainty on $N^\nu_\text{eff} $ to $0.01$~\cite{Abazajian:2013oma, Manzotti:2015ozr}. This would lead to a stronger lower bound on $m_\chi$ of about 12~MeV.

Note that  observations of supernova explosions only constrain small values of $\epsilon$ that are irrelevant here. In fact, kinetic mixing values larger than $\mathcal{O}(10^{-6})$ are enough to avoid that most of $Z'$ and DM particles escape the supernova core. Thus, the predictions of our scenario regarding supernovae are indistinguishable from those of the SM~\cite{Fradette:2014sza}.

\item{\textbf{Direct searches}}. The scattering of Majorana DM particles off nucleons is velocity-suppressed if such process is induced by the exchange of a $Z'$ boson coupled to a vector current of SM fermions. That is the case of the present scenario because, as said above, for $m_{Z' } \ll m_Z$,  the neutral current  $J_Z^\mu$ is approximately decoupled from the portal interactions. 
 Thus, this scenario can  be hardly constrained by current direct detection experiments.

\item {\textbf{Indirect searches}}.  As discussed at length previously, the p-wave annihilation channels responsible for the relic density are suppressed by at least two powers of the DM velocity. Moreover, other processes such as virtual internal bremsstrahlung or one-loop annihilations into photons are suppressed by the mass of the charged mediators that could induce them. 
One might think that for $m_{\chi} \ge m_{\pi}/2$, the processes $\chi\chi \to \pi^0 \gamma$ 
are relevant. However, they do not arise in the s-wave configuration as they require angular momentum $J=1$.  
Consequently, as already anticipated in Sec.~\ref{sec:SIDMtoSM}, this scenario can not be probed by indirect searches of DM.

\end{itemize}

\begin{figure*}[t]
\centering\includegraphics[width=3.2in]{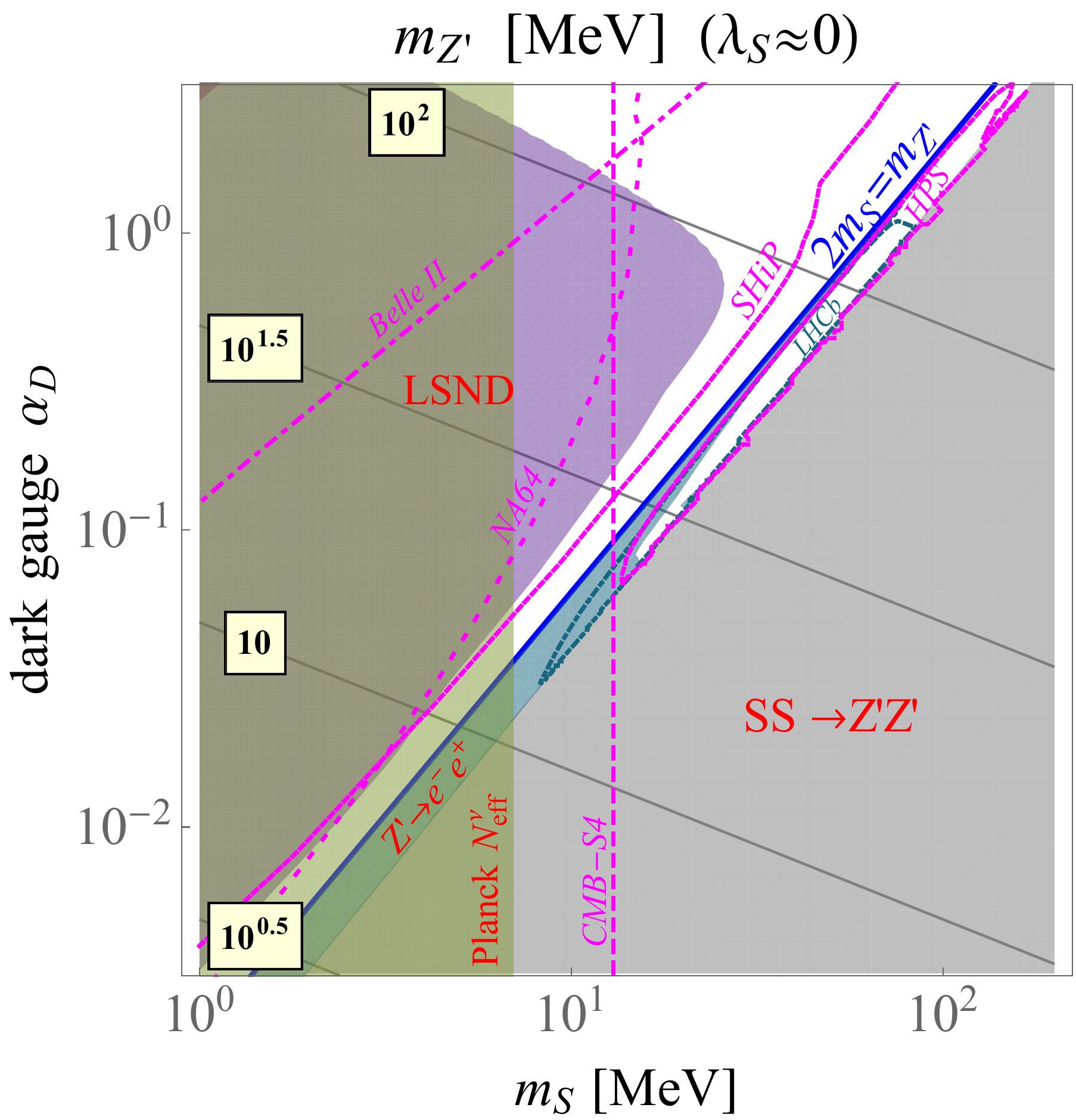}
\centering\includegraphics[width=3.2in]{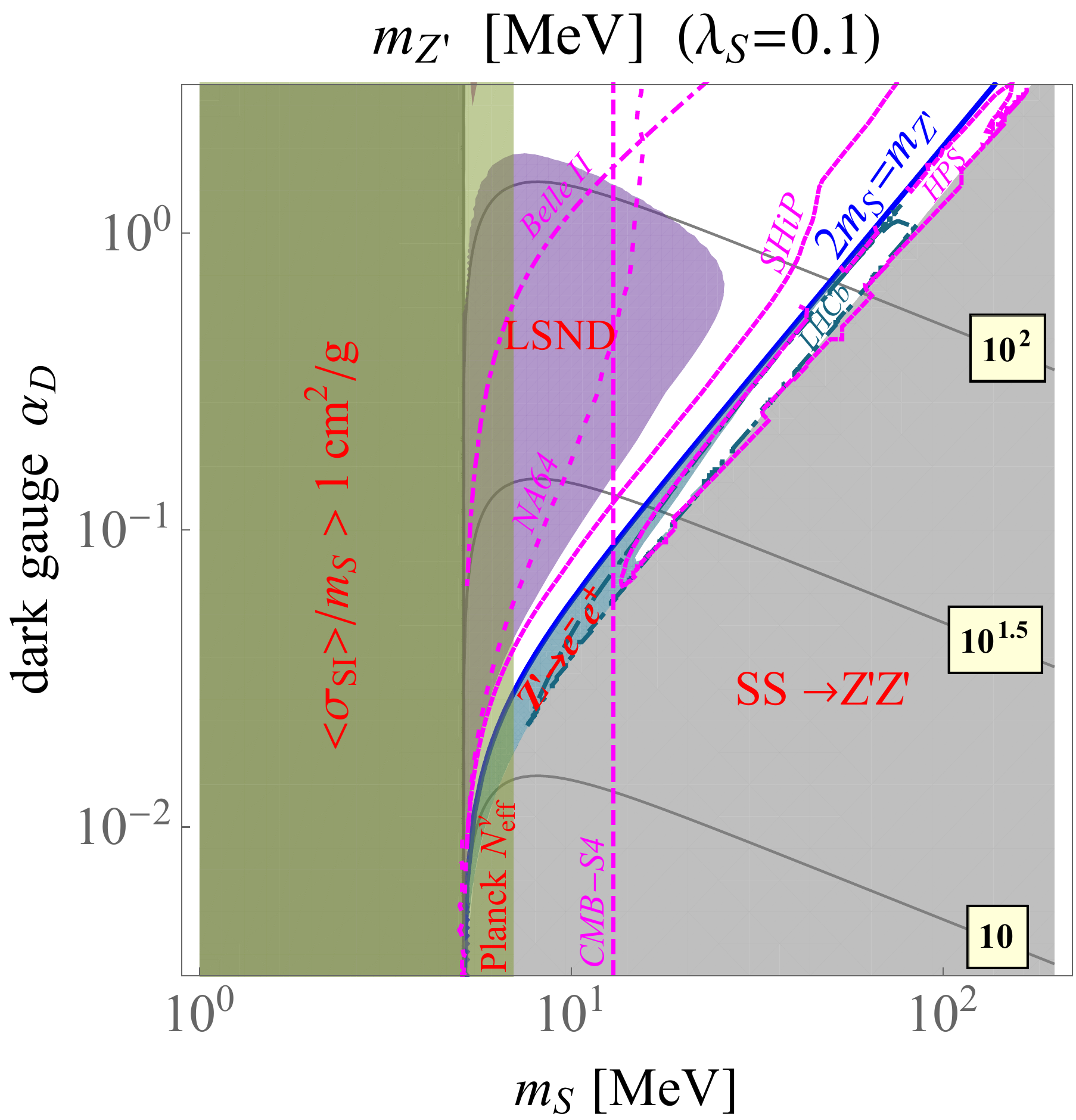}
\vskip -.5cm
\caption{
{\it $Z'$ portal for scalar DM.} 
As a function of the DM mass $m_S$ and dark coupling $\alpha_D$, the solid lines show the  $Z'$ mass satisfying the relic density and the self-interaction constraints, as given in Eqs.~\eqref{eq:Omegah2} and \eqref{Zportal:MajoranaSI}, for two choices of the scalar self-coupling  $\lambda_S$. Results are similar to the Majorana case above, specially   for $\lambda_S \sim 0 $. In the shaded region at \textit{right-bottom} corner, the dark freeze-out from $S S \rightarrow Z' Z'$ is too fast to account for the DM abundance. In the right panel, a lower bound $m_S \gtrsim 5.4$\,MeV holds due to the non-zero quartic coupling contribution to the self-interactions. Note that future direct detection experiments with semiconductor targets are expected  to probe all the allowed region~\cite{Essig:2015cda}.
}
\label{scalar:Zportal}
\end{figure*}

From this list of constraints, we conclude that Majorana DM coupled to a slightly heavier $Z'$ boson provides a viable model of self-interacting DM, that is
still allowed within a relatively large region of the parameter space. As shown in Fig.~\ref{fermion:Zportal}, the preferred DM masses lie around a few tens of MeV. While the HPS experiment and LHCb will probe a large fraction  of the  parameter space where the $Z'$ boson decays visibly, Belle-II  and the SHiP experiment at the CERN SPS will probe part of the region $2m_\chi \lesssim m_{Z'}$,  where it decays invisibly.

Before closing this section, we would like to comment on simple UV completions of this scenario. Since Majorana fermions can not carry any charge, their Z′ axial current can only arise from the spontaneous symmetry breaking of the $U(1)_D$ symmetry. For instance, the \emph{vev} of a single scalar $S$ with twice the $U(1)_D$ charge of the chiral DM fermion can induce both the Majorana mass of this  fermion and the mass of the $Z'$. Also, anomaly cancellation requires
extra fermions charged under $U(1)_D$, introduced either in a chiral way (with several extra Weyl fermions~\cite{Batra:2005rh}) or in a vector-like way (assuming a chiral partner for the DM field, at the price of allowing a new mass scale). The DM  can be lighter than all other hidden sector particles if its Yukawa coupling to $S$ is  relatively small with respect to the couplings determining the masses of $S$, $Z'$ and the extra fermions. This is always possible in the parameter space of our interest, as shown in Fig.~\ref{fermion:Zportal}.
In this case,  these additional particles do not change the phenomenology of interest in this work. The extra fermions decay into DM particles and their participation in the freeze-out is suppressed. Similarly, the scalar does not affect the freeze-out, and its contribution to DM self-interactions is suppressed by powers of the Yukawa coupling. Finally, the scalar $S$ must not strongly mix with the SM Higgs boson to  satisfy bounds from the Higgs  invisible decay and other DM searches.

\subsection{Scalar dark matter}

If DM is made of a scalar $S$ annihilating into SM particles via a s-channel exchange of $Z'$ bosons, 
one obtains a scenario similar to the Majorana case, except for three important differences: DM is not its own antiparticle\footnote{For the sake of simplicity, we assume DM to be symmetric, i.e. the abundance of $S$ and $S^*$ are taken equal.}, there is an extra source of self-interaction for the DM and the direct detection rate is not anymore velocity suppressed.

First of all, let us look at the annihilation process, which is induced by the current $J_\text{DM}^\mu = i (S^* \partial_\mu S - S \partial_\mu S^*)$. In the  non-relativistic limit, the corresponding  cross section is given by
\be
\sva   \simeq \frac{4\pi \epsilon^2\alpha \alpha_D  \sum_{f} \,N_f\, q_f^2 \,   (1-r_f)^{1/2} (2+r_f)\,v^2}{3 m_{S}^2\left((r_{Z'} - 4)^2 + r_{Z'}^2\Gamma_{Z'}^2/{m^2_{Z'}}\right)
}  \,,
\ee
Summing over all kinematically allowed channels in the same way as for the Majorana case above, Eq.~\eqref{eq:Omegah2} fixes the DM relic density.

As for the DM self-interactions, they are induced by  the exchange of the $Z'$ boson, and possibly by an additional  $ {\cal L_S} \supset -\lambda_S (S^* S)^2$ quartic coupling contribution. Due to the fact that DM is not is own antiparticle, there are several self-interaction channels, namely, $SS\leftrightarrow SS$, $SS^*\leftrightarrow S S^*$ and $S^*S^ *\leftrightarrow S^*S^*$.  
The corresponding averaged cross section in the non-relativistic limit is
\be
{ \ssi \over m_S} =  \frac{3\lambda_S^2}{16\pi\, m_{S}^3}+  \frac{6 \pi\,\alpha_D^2 m_S}{m_{Z'}^4}
\ee
For the case where one assumes a negligible value of the quartic coupling, the left panel of  Fig.~\ref{scalar:Zportal} shows, as a function of $m_{S}$ and $\alpha_D$, the values of $m_{Z'}$ that one needs in order to fulfill both the relic density constraint and the benchmark $\ssi /m_\text{DM}=\unit[1]{cm^2/g}$.
Likewise,  the right panel of  Fig.~\ref{scalar:Zportal} shows the corresponding situation when we switch on the scalar coupling, taking  $\lambda_S=0.1$ as a sample value. 

In the latter case, as shown in the right panel of Fig.~\ref{scalar:Zportal}, the self-interaction hypothesis precludes too light dark matter candidates independently of the value of $\alpha_D$,  because the scalar coupling contribution to the self-interaction cross section scales as $1/m_S^2$.
Also, note that having a large value of $\lambda_S$ at such a low scale may give rise to a Landau pole below the electroweak scale  (unless there are extra low energy degrees of freedom in the hidden sector contributing negatively to the $\beta$ function of this coupling
\footnote{
Notice that a Landau pole can also develop for the  scalar $S$  introduced in the UV completion of the Majorana scenario above. Nevertheless,
such Landau pole can be easily avoided  because there the heavier fermions introduced for anomaly cancellation 
provide such negative contribution.
}). For example in a pure $\lambda_S S^4$ theory, a value $\lambda_S=0.1$ at $m_{S}\sim10-100$~MeV scale leads to a Landau pole around the electroweak scale.

As mentioned at the beginning of this section, unlike for Majorana particles, the scalar case does not lead to velocity-suppressed direct detection cross sections. 
Although currently sub-GeV DM is almost  unconstrained by direct detection experiments, 
such an unsuppressed  rate may lead to potential tests in the future. Experiments searching for nuclear recoil are not so promising in this regard. For example, the most optimistic case for our purpose is the future SNOLAB experiment which will be able to probe DM particles with masses down to $\unit[0.5]{GeV}$.   However, a signal could be seen in experiments searching for DM-electron collision. For scalar DM communicating with the SM via a $Z'$-portal, this has been studied in for XENON10~\cite{Essig:2011nj, Essig:2012yx} . Here, the collision cross section is given by $\sigma_e \sim {16  \pi \epsilon^2 \alpha_D \alpha m_\text{e}^2 / m_{Z'}^4 }$. Lastly, future experiments with semiconductor targets are expected to be able to probe the whole parameter region allowed today~\cite{Essig:2015cda}.
\vspace{2cm}

\section{Conclusions}
\label{sec:conclusions}

In this work, we have shown that it is possible for a
self-scattering DM particle (with a strength capable
of addressing the small scale structure problems of the CDM paradigm) to freeze out  dominantly from annihilations into SM particles.

We have argued that this is only possible
if the DM mass  lies below the GeV scale. Barring large fine-tuning, this immediately implies that DM must be a singlet of the SM gauge group. The same remark applies for any particle mediating the annihilation  process, because otherwise such mediator would need to be around the electroweak scale or above,  and the corresponding annihilation rates would be  suppressed. These facts together imply that the DM and the SM sectors must be connected through one or several of the four SM singlet portal interactions, associated to a scalar boson, a right-handed neutrino and a $Z'$ massive gauge boson. 

We  have shown that  only the option of a $Z'$ boson coupled to Majorana or scalar DM passes all the experimental constraints. From its simplicity and the fact that it does not require any special tuning, this scenario constitutes an 
attractive way to accommodate both DM large self-interactions and the relic density constraint. Here, the huge difference between the self-interaction and annihilation cross sections is not due to any special mechanism taking place; it is simply due to the fact that the portal interaction, which enters in the annihilation but not in the self-interaction, is suppressed. Furthermore, this scenario offers possibilities of particle physics tests.

Quantitatively, Fig.~\ref{fermion:Zportal} (for the Majorana case) and Fig.~\ref{scalar:Zportal} (for the scalar case) summarize the various constraints and future possibilities of testing it or ruling it out.  For the scalar case, in addition to the constraints shown in Fig.~\ref{scalar:Zportal},  semiconductor target direct detection experiments have the potential to probe all the parameter space allowed today.

\section*{Acknowledgments}
We thank Julian Heeck and Josh Ruderman for useful discussions.
The work of C.G.C. and T.H. is supported by the FNRS, by the IISN and by the 
Belgian Federal Science Policy through the Interuniversity Attraction Pole P7/37. 

\newpage
\bibliographystyle{apsrev}
\bibliography{text}

\end{document}